\UseRawInputEncoding
\documentclass[aps,prb,reprint,superscriptaddress,compress]{revtex4-2}
\usepackage{graphicx}
\usepackage{dcolumn}
\usepackage{bm}
\usepackage{tabularx}
\usepackage[T1]{fontenc}
\usepackage{booktabs,array,mathptmx,tabularx,booktabs,lipsum,amsmath,multirow}
\usepackage{endnotes}

\makeatletter

\newcommand{\Rmnum}[1]{\expandafter\@slowromancap\romannumeral #1@}
\makeatother

\emergencystretch=\maxdimen
\begin{document}

\title{Pressure-tuned superconductivity in the Dirac semimetal PdTe}

\author{Xiangming Kong}
\thanks{These authors contributed equally to this work.}
\affiliation{State Key Laboratory of Surface Physics, Department of Physics, Fudan University, Shanghai 200438, China}
\affiliation{Shanghai Research Center for Quantum Sciences, Shanghai 201315, China}

\author{Zhaolong Liu}
\thanks{These authors contributed equally to this work.}
\affiliation{Beijing National Laboratory for Condensed Matter Physics, Institute of Physics, Chinese Academy of Sciences, Beijing 100190, China}

\author{Xiangqi Liu}
\thanks{These authors contributed equally to this work.}
\affiliation{School of Physical Science and Technology, ShanghaiTech University, Shanghai 201210, China}

\author{Chunqiang Xu}
\affiliation{School of Physical Science and Technology, Ningbo University, Ningbo 315211, China}

\author{Zhenhai Yu}
\affiliation{School of Physical Science and Technology, ShanghaiTech University, Shanghai 201210, China}

\author{Jing Wang}
\affiliation{State Key Laboratory of Surface Physics, Department of Physics, Fudan University, Shanghai 200438, China}

\author{Baomin Wang}
\affiliation{School of Physical Science and Technology, Ningbo University, Ningbo 315211, China}

\author{Xiaofeng Xu}
\affiliation{Department of Applied Physics, Zhejiang University of Technology, Hangzhou 310023, China}

\author{Yanfeng Guo}
\affiliation{School of Physical Science and Technology, ShanghaiTech University, Shanghai 201210, China}
\affiliation{ShanghaiTech Laboratory for Topological Physics, Shanghai 201210, China}

\author{Rui Zhang}
\email{zhangr4@tcd.ie}
\affiliation{Northwest institute for non-ferrous metal research, Xi'an 710016, China}

\author{Xiaofan Yang}
\email{yangxiaofan@fudan.edu.cn}
\affiliation{State Key Laboratory of Surface Physics, Department of Physics, Fudan University, Shanghai 200438, China}
\affiliation{Northwest institute for non-ferrous metal research, Xi'an 710016, China}

\author{Shiyan Li}
\email{shiyan$\_$li$@$fudan.edu.cn}
\affiliation{State Key Laboratory of Surface Physics, Department of Physics, Fudan University, Shanghai 200438, China}
\affiliation{Shanghai Research Center for Quantum Sciences, Shanghai 201315, China}
\affiliation{Collaborative Innovation Center of Advanced Microstructures, Nanjing 210093, China}

\date{\today}

\begin{abstract}
We report an unusual evolution of superconductivity (SC) in the three-dimensional Dirac semimetal PdTe with increasing pressure up to $\sim$50 GPa. The compressed PdTe exhibits a sudden reversal in the superconducting transition temperature $T\mathrm{_c}$, from an initial decrease with pressure to an increase above a critical pressure $P\mathrm{_c} \approx$ 15 GPa, showing a V-shaped feature composed of SC-\Rmnum{1} and SC-\Rmnum{2} phases due to a structural phase transition through $P\mathrm{_c}$. Subsequently, the $T\mathrm{_c}$ goes into a plateau around 2.5 K when the pressure is higher than $\sim$32 GPa. In addition, we find the variations of carrier concentrations and mobilities also manifest a similar trend on the pressure response as the $T\mathrm{_c}$, and the normal-state electronic properties change from the electron-dominated single-band model to two-carrier model after the structural phase transition, implying the close correlation between electronic properties and two SC phases. Our findings establish the SC of PdTe is highly tunable under varying pressures.
\end{abstract}
\maketitle

Transition metal dichalcogenides (TMDCs) have become a widely accepted platform to investigate novel quantum states because of the topological nontrivial nature of the electronic band structure. For instance, PdTe$_2$, an impressive example in palladium tellurides, has been classified as a type-II Dirac semimetal which implies a tilted Dirac cone with a topologically nontrivial surface state \cite{1_soluyanov_type-ii_2015,2_bahramy_ubiquitous_2018,3_clark_fermiology_2018,4_fei_nontrivial_2017,5_amit_type-ii_2018,6_cook_observation_2023}. Moreover, the coexistence of topological electronic states and superconductivity \cite{7_leng_type-i_2017,8_kim_importance_2018,9_das_conventional_2018,10_leng_type-i_2019,11_anemone_electronphonon_2021,12_liu_two-dimensional_2018,13_liu_type-ii_2020} provides an opportunity for exploration of the interplay between superconducting properties and Dirac fermions. PdTe$_2$ forms through the stacking of layers containing PdTe$_6$ octahedron. A similar characteristic shares with its monochalcogenide counterpart PdTe, considering as a potential topological material with superconductivity (SC). However, substantial progress in PdTe has been reported only recently due to the improvements in synthesis methods \cite{14_yang_coexistence_2023,15_wang_ionic_2023,16_wu_synthesis_2022}.\par

In contrast to the conventional superconductor PdTe$_2$ \cite{7_leng_type-i_2017,8_kim_importance_2018,9_das_conventional_2018,10_leng_type-i_2019,11_anemone_electronphonon_2021}, whether PdTe has unconventional SC is still controversial. Recent angle-resolved photoemission spectroscopy (ARPES) firstly reports PdTe is a three-dimensional (3D) Dirac semimetal (DSM) with the coexistence of the bulk-nodal and surface-nodeless SC in PdTe \cite{14_yang_coexistence_2023}. Subsequent specific heat and quantum oscillation measurements also suggest the potential of PdTe as an unconventional and topological superconductor \cite{17_chapai_evidence_2023}. However, the most recent ultralow-temperature thermal conductivity measurements reveal a multiband nodeless s-wave superconducting gap structure \cite{18_zhao_multigap_2023}. Overall, these contradictory results show that the superconducting mechanism of PdTe is debatable. Further exploration of additional external control parameters, such as pressure, is warranted to gain new insights into the superconducting behavior of PdTe.\par

Pressure provides a practical way to tune crystal and electronic structures without introducing any chemical impurities. By manipulating the interplay between charge, spin, orbital and lattice degrees of freedom, pressure has been widely employed to tailor SC behavior in various superconductors \cite{19_zhou_pressure-induced_2016,20_dong_structural_2021,21_tafti_sudden_2013,22_mukasa_high-pressure_2021,23_zhou_quantum_2022,24_zheng_emergent_2022,25_sun_signatures_2023}. And most encouragingly, the effect of pressure on the superconducting transition temperature ($T\mathrm{_c}$) of PdTe$_2$ shows a nonmonotonic variation without structural phase transition (SPT)  \cite{26_leng_superconductivity_2020,27_furue_superconducting_2021,28_soulard_why_2005}. The related analysis reveals that the evolution of $T\mathrm{_c}$ in PdTe$_2$ under pressure depends on the competition between the electronic density of states and lattice stiffening. As a result, we anticipate that pressure can also induce some novel behaviors of SC in PdTe based on the same PdTe$_6$ octahedral unit.\par

In this Letter, high-pressure transport and synchrotron diffraction measurements have been performed to explore SC in PdTe single crystal. Initially $T\mathrm{_c}$ of PdTe decrease from 4.5 K to a minimum of $\sim$1.9 K at the critical pressure $P\mathrm{_c} \approx$ 15 GPa and it then gradually goes up to $\sim$2.5 K, showing a V-shaped temperature-pressure curve. The Hall resistivity exhibits the transition from a single-band model to a two-carrier model through $P\mathrm{_c}$, showing a consistent behavior between the normal-state electronic properties and superconducting state over the entire pressure range. Both theoretical calculations and the structural analysis indicate the pressurized PdTe exists a structural phase transition through $P\mathrm{_c}$, and corresponding discussion certifies the origins in SC-\Rmnum{1} and SC-\Rmnum{2} phases distinguished by SPT are closely associated with the electronic properties, particularly the contribution of Pd $d$-orbital electron and Te $P$-orbital hole is important to the increase of $T\mathrm{_c}$ in SC-\Rmnum{2} phase.\par
\begin{figure}[htb]
	\centering
	\includegraphics[clip,width=0.45\textwidth]{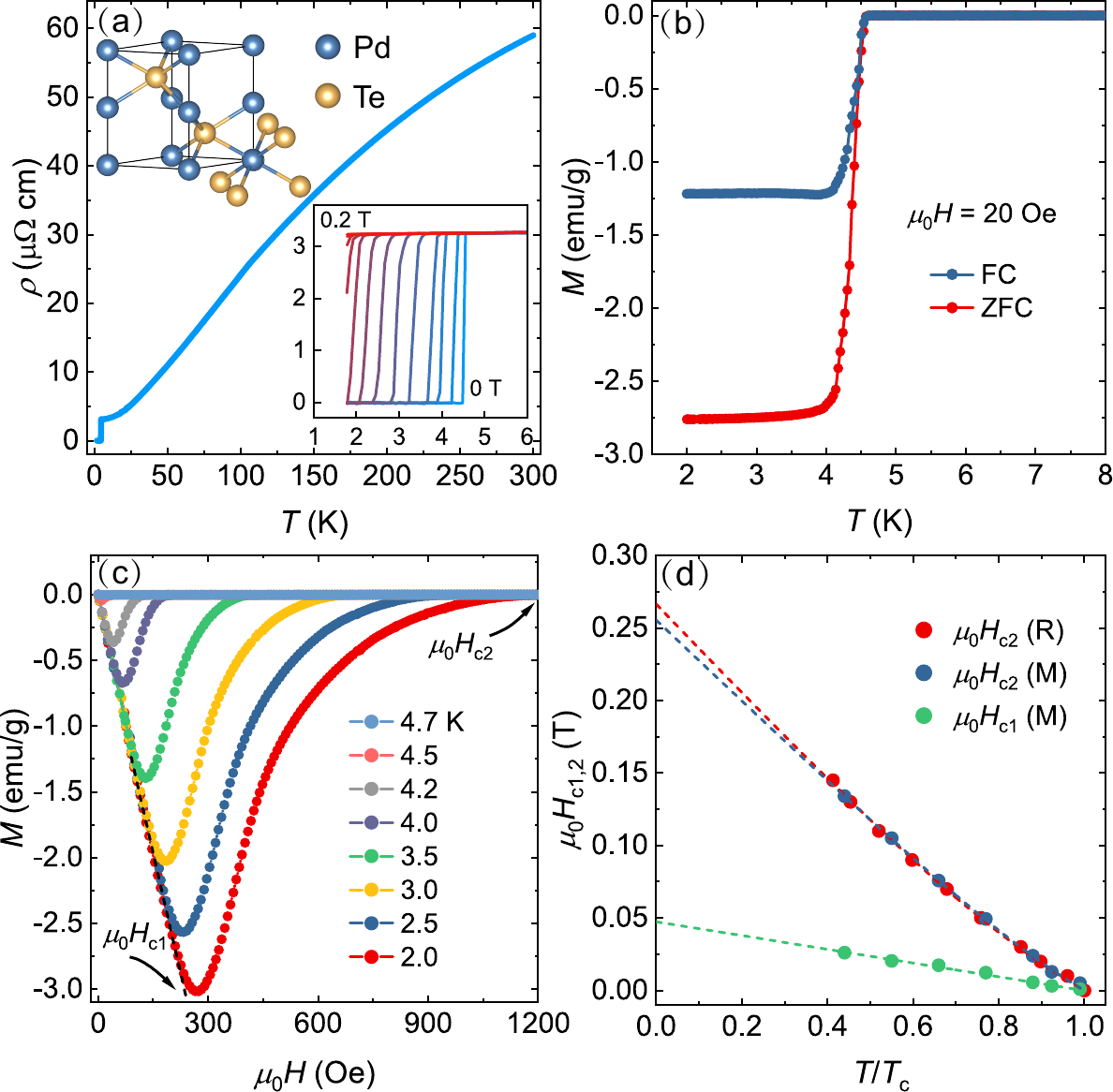}
	\caption{(a) Temperature dependence of resistivity of PdTe single crystal. $T\mathrm{_c}$ is defined at 90\% of the normal state resistivity. The insets show crystal structure and PdTe$_6$ octahedral unit (top left) and resistivity of PdTe under out-of-plane magnetic fields (bottom right). (b) DC magnetization measured in zero-field cooled (ZFC) and field cooled (FC) conditions for PdTe with 20 Oe applied magnetic field. (c) Isothermal magnetizations in the superconducting state of PdTe as a function of the applied out-of-plane magnetic field. (d) Extracted lower magnetic critical field $\mu\mathrm{_0}H\mathrm{_{c1}}$ and upper magnetic critical field $\mu\mathrm{_0}H\mathrm{_{c2}}$ from inset of (a) and (c) as a function of temperature. The dashed curves are fitted by the two-band model.}
\end{figure}

Single crystals of PdTe were grown using a self-flux method \cite{14_yang_coexistence_2023,17_chapai_evidence_2023}. Electronic transport measurements were conducted in a physical property measurement system (PPMS, Quantum Design), while  magnetization measurements were carried out by a magnetic property measurement system (MPMS3, Quantum Design). High-pressure conditions were generated using a nonmagnetic diamond anvil cell (DAC) with platinum foil electrodes, and a BeCu plate as gasket covered by a cubic BN/epoxy mixture insulator layer. High-pressure resistivity and Hall measurements were conducted by the standard four-probe method under van der Pauw configuration. In situ high-pressure X-ray diffraction (XRD) experiments were performed at the BL15U1 beamline ($\lambda$ = 0.6199 \AA) of the Shanghai Synchrotron Radiation Facility. Diffraction rings were then integrated into powder diffraction patterns using the Dioptas software. Pressure-dependent crystal structures and lattice parameters were analyzed, indexed, and refined by Jade Pro 8.8, and Fullprof software. Pressure in the DAC was determined by the ruby luminescence method \cite{29_mao_calibration_1986}. The first-principles calculations were carried out with density functional theory (DFT) implemented in the Vienna ab initio simulation package (VASP) \cite{30_kresse_efficiency_1996}, adopting the generalized gradient approximation (GGA) in the form of the Perdew-Burke-Ernzerhof (PBE) \cite{31_perdew_generalized_1996} for the exchange correlation potentials. Projector-augmented-wave (PAW) \cite{32_kresse_ultrasoft_1999} pseudopotentials were used with a plane wave energy of 600 eV. $4d^{10}$ of Pd and $5s^{25}p^4$ of Te electronic configurations were treated as valence electrons respectively. A Monkhorst-Pack Brillouin zone sampling grid \cite{33_monkhorst_special_1976} with a resolution of $0.02 \times 2\pi$ \AA$^{-1}$ was applied. Atomic positions and lattice parameters were relaxed until all the forces on the ions were less than $10^{-3}$ eV/\AA. The spin-orbit coupling (SOC) effect was included.\par
\begin{figure*}[htb]
	\centering
	\includegraphics[clip,width=0.8\textwidth]{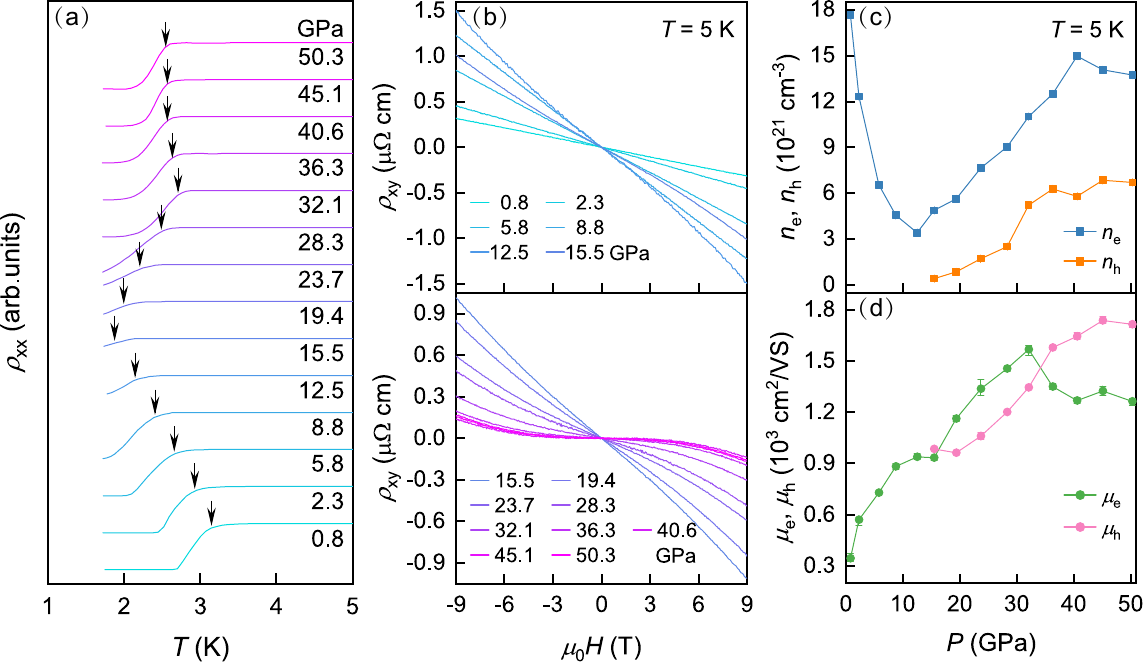}
	\caption{(a) Temperature-dependent resistivity of PdTe under pressure. Black arrows indicate the $T\mathrm{_c}$. (b), (c) The Hall resistivity of PdTe with the magnetic field applied along the $c$ axis at 5 K. The Hall resistivity exhibits a typical two-carrier behavior above the critical pressure $P\mathrm{_c} \approx$ 15 GPa. (d), (e) Extracted carrier concentrations and mobilities in the one-band and two-carrier regimes.}
\end{figure*}
\begin{figure*}[htb]
	\centering
	\includegraphics[clip,width=0.8\textwidth]{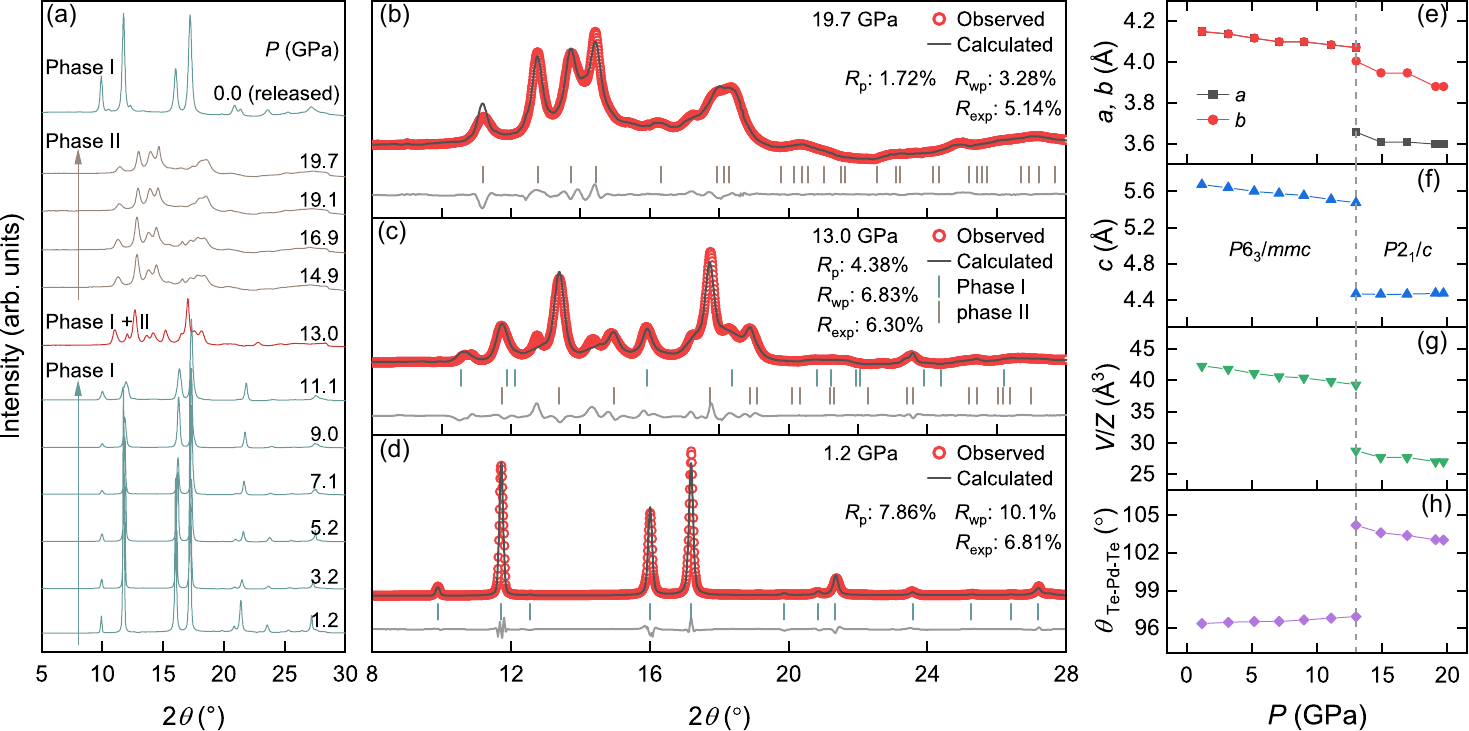}
	\caption{(a) Representative powder synchrotron X-ray diffraction patterns at room temperature ($\lambda$ = 0.6199 \AA). (b)-(d) Three Rietveld refinements profiles of PdTe at 19.7 GPa, 13.0 GPa and 1.2 GPa. (e), (f) Pressure-dependent lattice parameters $a$, $b$ and $c$ for the pristine hexagonal ($P6_3/mmc$, $Z$ = 2) and high-pressure monoclinic ($P2_1/c$, $Z$ = 2) phases extracted from refinements. (g) Pressure-dependent unit-cell volume. (h) Pressure-dependent the angle formed between Te-Pd-Te in  a PdTe$_6$ octahedron}
\end{figure*}

PdTe crystallizes in the NiAs-type hexagonal structure with space group $P6_{3}/mmc$ (No.194) with the lattice parameters $a = b = 4.152$ \AA\ and $c = 5.671$ \AA. The 3D structure of PdTe comprises edge-shared PdTe$\mathrm{_6}$ octahedron (inset of Fig. 1(a)). Given its structural consistency with FeTe, early studies into PdTe have primarily focused on comparison and extension of the understanding of unconventional SC in iron chalcogenides \cite{34_karki_pdte_2012,35_karki_interplay_2013,36_ekuma_first-principles_2013}. PdTe shows a superconductivity with a critical temperature $T{\mathrm{_c}}$ of 4.5 K (defined at 90\% of the normal-state resistivity) under ambient conditions. The $T\mathrm{_c}$ moves to lower temperatures with increasing magnetic fields up to $\sim$0.25 T (inset of Fig. 1(a)). A clear diamagnetic response below the $T\mathrm{_c}$ in both zero-field cooled (ZFC) and field-cooled (FC) magnetization modes indicate that the resistivity drop corresponds to the superconducting transition (Fig. 1(b)). The isothermal magnetization ranging from 2 to 4.7 K suggests the nature of type-\Rmnum{2} SC in PdTe (Fig. 1(c)). Moreover, it is noted that the normalized temperature dependence of the lower ($\mu\mathrm{_0}H\mathrm{_{c1}}$) and upper critical field ($\mu\mathrm{_0}H\mathrm{_{c2}}$) extracted from the isothermal magnetization and magneto-resistivity measurements (Fig. 1(d)) satisfy well with the two-band model \cite{37_gurevich_enhancement_2003}. This feature agrees with the multiband concepts concluded from the superconducting specific heat and ultralow-temperature thermal conductivity measurements \cite{17_chapai_evidence_2023,18_zhao_multigap_2023}. \par
\begin{figure*}[htb]
	\centering
	\includegraphics[clip,width=0.8\textwidth]{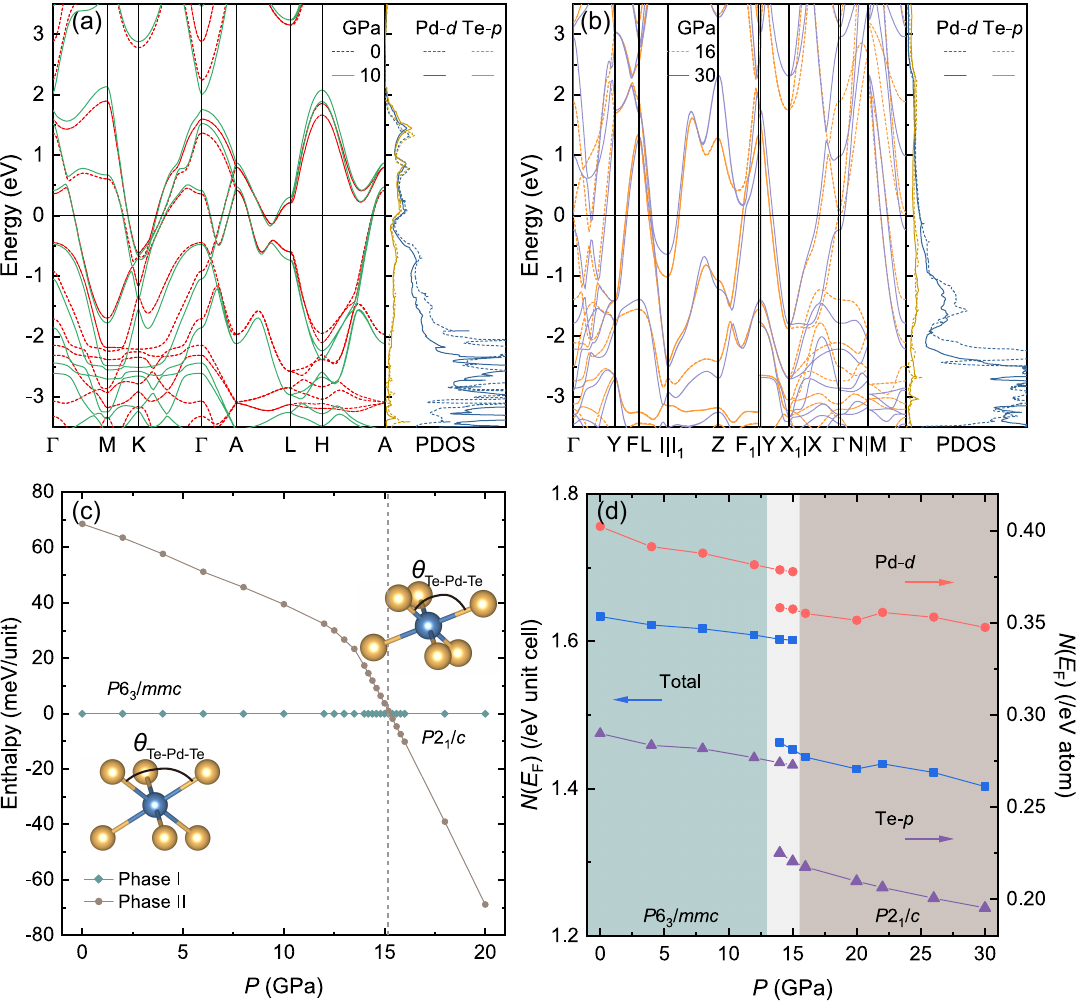}
	\caption{Electronic band structure calculations of PdTe (a) before and (b) after the SPT. In each figure, the left panel shows the projected band structures of PdTe with spin-orbit coupling included at various pressures, and the right panel presents the projected electronic density of states (PDOS). The black line marks the Fermi level. (c) Relative enthalpy of PdTe as a function of pressure. The insets depict the definition of bond angle $\theta\mathrm{_{Te-Pd-Te}}$ in a PdTe$_6$ octahedron. The enthalpy of the phase \Rmnum{1} is taken as a reference. (d) Density of states at the Fermi level $N(E\mathrm{_F})$ as a function of pressure.}
\end{figure*}

Figure 2(a) shows the temperature dependence of the longitudinal resistivity $\rho\mathrm{_{xx}}$ for the PdTe single crystal under various pressures ranging from 0.8 to 50.3 GPa. The SC spans over the whole pressure range. The $T\mathrm{_c}$ initially drops from 3.1 K to a minimum of $\sim$1.9 K, followed by an increase to 2.7 K as pressure varies. The critical pressure $P\mathrm{_c}$ meets the minimum $T\mathrm{_c}$ at $\sim$15 GPa. In addition, when the pressure exceeds 32.1 GPa, a plateau of $T\mathrm{_c}$ with $\sim$2.5 K is observed. Earlier first-principles calculations support that the initial decrease of $T\mathrm{_c}$ origins from the combined effects of the phonon hardening and the reduced electron-phonon coupling \cite{38_chen_superconductivity_2016,39_cao_elastic_2015}. However, the abnormal increase in $T\mathrm{_c}$ above $P\mathrm{_c}$ has not been reported. Figures 2(b)-(c) display the Hall resistivity $\rho\mathrm{_{xy}}$ near superconducting state ($T$ = 5 K) under the corresponding pressure in Fig. 2(a). In contrast to an electron-dominated single-band behavior below $P\mathrm{_c}$, the $\rho\mathrm{_{xy}}$ above $P\mathrm{_c}$ can be divided into two parts (linear behavior in the high-field region and “S” shape in the low-field region), indicating that electron and hole coexist. In general, the significant changes in Hall resistivity implies alterations in the band structure. By fitting the $\rho\mathrm{_{xy}}(\mu_{0}H)$ with the single-band model and two-carrier model at various pressures, the carrier concentration and mobility can be extracted in Figs. 2(d)-(e). Below the $P\mathrm{_c}$, electron concentration reduces while electron mobility gradually increases. When the pressure exceeds 15 GPa, both electron and hole concentrations monotonically increase, and then the net carrier concentration tends to level off, the trend of which is similar to the pressure-dependent $T\mathrm{_c}$ evolution. The high correlation between the electronic properties and the superconducting state indicates that a transition in the band structure plays a crucial role in manipulating the SC evolution.\par
\begin{figure}[htb]
	\includegraphics[clip,width=0.4\textwidth]{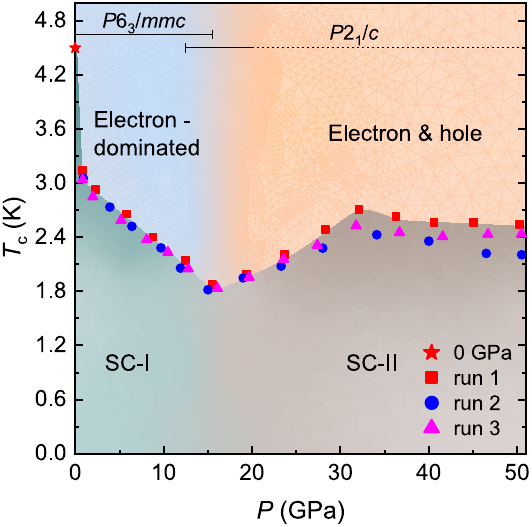}
	\caption{Temperature-pressure phase diagram of PdTe. SC-\Rmnum{1} and SC-\Rmnum{2} stand for the two superconducting phases of PdTe under pressure. Three independent runs were performed.}
\end{figure}

To investigate whether the enhanced SC and band feature in PdTe are associated with a structural phase transition (SPT) or an amorphous phase transition, in situ high pressure synchrotron diffractions on pulverized PdTe samples were conducted with the pressure ranging from 1.2 to 19.7 GPa, as shown in Fig. 3(a). Under lower pressures ($\leq$ 11.1 GPa), the XRD patterns can be accurately indexed using hexagonal crystallography (space group: $P6_{3}/mmc$, phase \Rmnum{1}). Notably, all representative diffraction peaks undergo a shift towards higher angles, and no new peaks are observed. Upon reaching 13.3 GPa, the appearance of unexpected diffraction peaks beyond primitive phase, suggesting a SPT. Under 19.7 GPa, a distinctive pattern emerging signifies that PdTe forms a monoclinic structure (phase \Rmnum{2}) with space group $P2_{1}/c$ (No.14), which is consistent with the Rietveld refinements, as indicated in Fig. 3(b). At 13.3 GPa, the two-phase refinement suggests the coexistence of phase \Rmnum{1} and \Rmnum{2}. It is noted that the XRD pattern in the released state is almost identical to the initial one, demonstrating the reversibility of the SPT. As depicted in Figs. 3(e)-(g), unlike the synchronous contraction of lattice parameters $a$, $b$ and $c$ before the SPT, three distinguished compression behaviors emerge after the SPT: lattice parameters $a$ and $b$ undergo varying degrees of decrease, while $c$ keeps essentially constant. This drives to a piecewise continuous reduction in the unit cell volume $V/Z$ through the $P\mathrm{_c}$ where $Z$ is the number of asymmetry units in each crystal lattice. Typically, pressure-induced symmetry changes may give rise to Fermi surface reconstructions, and corresponding transports anomalies, as evidenced by our findings in electrical resistivity. Meanwhile, the pressure dependence of the Te-Pd-Te angle $\theta\mathrm{_{Te-Pd-Te}}$ (definition is given in inset of Fig. 4(c))  present earlier dilative and later contractive trend. The high-pressure investigations of SC in PdTe$_2$ suggest a direct correlation between the nonmonotonic variation of $T\mathrm{_c}$ and $\theta\mathrm{_{Te-Pd-Te}}$ \cite{26_leng_superconductivity_2020,27_furue_superconducting_2021}. Ref.  \cite{27_furue_superconducting_2021} argues that the variation of $\theta\mathrm{_{Te-Pd-Te}}$ results in varying phonon frequencies, thereby modulating the SC of PdTe$_2$. Hence, the examination of electronic band structures and crystal phases offers two distinct perspectives for investigating the effects of pressure on SC in PdTe.\par

Electronic band structure calculations (Figs. 4(a)-(b)) show noticeable pressure-induced band structure changes as we expected. The band structure remains largely unchanged around the Fermi level ($E\mathrm{_F}$) at 10 GPa compared to that at ambient pressure conditions. However, a completely different band structure shows up after the SPT. This transition is in agreement with the significant changes in both transport and diffraction experiments. Furthermore, increased pressures drive the $E\mathrm{_F}$ shift to lower levels, which also can be reflected in the lower energy distribution of each orbital projected density of states. Figure 4(d) illustrates the evolution of DOS at $E\mathrm{_F}$ ($N(E\mathrm{_F})$) for phase \Rmnum{1} and phase \Rmnum{2} in respect to pressure. A slight decrease of $N(E\mathrm{_F})$ in phase \Rmnum{1} is consistent with the existing theoretical analysis \cite{37_gurevich_enhancement_2003,38_chen_superconductivity_2016}. However, the $N(E\mathrm{_F})$ from the reconstructed band structure alone is not sufficient to judge the trend in $T\mathrm{_c}$ change, suggesting that the phonon- and electron-phonon interactions are crucial to enhance the $T\mathrm{_c}$ of PdTe within the SPT according to the Bardeen-Cooper-Schrieffer (BCS) theory \cite{40_bardeen_theory_1957,41_allen_neutron_1972}. This inference is also be used to explain why Pd$_{1-x}$Ni$_x$Te does not exhibit SC when $x$ > 0.2 despite $N(E\mathrm{_F})$ increases monotonically with Ni concentration \cite{42_kumar_first-principles_2020}. Moreover, note that both the electron concentration (as shown in Figs. 2(d)-(e)) and $N(E\mathrm{_F})$ are synchronized with $T\mathrm{_c}$ of PdTe with incerasing pressure before the SPT, indicating the normal-state (near superconducting state) electronic properties can be regarded as an intuitive indicator to assess the $N(E\mathrm{_F})$. However, the consistency is violated after the SPT, showing the opposite trends in the $N(E\mathrm{_F})$ and carrier (electron and hole) concentrations. A simple but plausible explanation for this could be that our calculations severely underestimate the contribution of the Pd $d$-orbital and Te $p$-orbital to $N(E\mathrm{_F})$. From orbital decomposition analysis in Ref. \cite{17_chapai_evidence_2023}, the electronic and hole natured Fermi surfaces are dominated by Pd $d$-orbital and Te $p$-orbital, respectively. Therefore, the increasing electron concentration and mobility and unexpected hole feature after the SPT (as shown in Figs. 2(d)-(e)) can significantly associated both with Pd $d$-orbital and Te $p$-orbital. By contrast, our calculated $N(E\mathrm{_F})$ contributed by Pd $d$-orbital and Te $p$-orbital in Fig. 4(d) is always decreasing, resulting in the reduction of total $N(E\mathrm{_F})$. Additionally, we calculated the pressure-dependent enthalpy of phase \Rmnum{2} with respect to that of phase \Rmnum{1} from 0 to 20 GPa as seen in Fig. 4(c). The phase \Rmnum{2} acquires the ascendancy in energy at 15.2 GPa. This characteristic pressure is in consistent with the critical pressure $P\mathrm{_c}$ identified by transport and synchrotron diffraction experiments.\par

Figure 5 summarizes the temperature-pressure ($T-P$) phase diagram of PdTe. There is an abrupt decrease of $T\mathrm{_c}$ at $\sim$1 GPa in SC-\Rmnum{1} phase region, contrasting sharply with that under the ambient pressure at 4.5 K, even though the structural parameters have changed only slightly (the compression along the $a$- and $b$-axis directions is less than 0.12\% at 1.2 GPa, as shown in Figs. 3(e)-(g)). A similar phenomenon has also been seen in Pd$_{1-x}$Ni$_x$Te, PdTe$_{1-x}$Sb$_x$ and Pd$_{1-y}$Te \cite{43_goyal_impact_2016,44_chen_hole_2021}, where introduced elements, such as Ni, Sb and Pd vacancy, act as chemical pressure, leading to a decrease in $T\mathrm{_c}$. Ref. \cite{44_chen_hole_2021} suggests the variations of $\theta\mathrm{_{Te-Pd-Te}}$ can reflect the contribution of phonon mode for SC. The direct evidence is that the $T\mathrm{_c}$ of the above mentioned doped PdTe samples is all sensitive to change in $\theta\mathrm{_{Te-Pd-Te}}$, such as, a change of $0.1^{\circ}$ can result in a decrease from 4.5 K to $\sim$3 K. We note that the influence of $\theta\mathrm{_{Te-Pd-Te}}$ on the SC of PdTe under physical pressure in Fig. 3(g) and Fig. 5 before the SPT seems more akin to the isoelectronic Ni doping (with a greater band angle leading to a lower $T\mathrm{_c}$), while Sb doping and Pd deficiency can induce a distinct modulation of charge carriers. Combining our DFT calculations and related discussion, it can be inferred that the $N(E\mathrm{_F})$ and phonon mode dominate the evolution of SC in PdTe before the SPT. Additionally, the dilation of $\theta\mathrm{_{Te-Pd-Te}}$ might be an index of phonon hardening.\par

As for SC-\Rmnum{2} phase region, the mechanism of $T\mathrm{_c}(P)$ becomes more complex. According to McMillan formula \cite{45_mcmillan_transition_1968,46_phan_kohn-luttinger_2020}, one route to the increasing $T\mathrm{_c}$ is to increase phonon frequency. However, the phonon frequency is tightly coupled to the electron-phonon coupling constant (which varies inversely as the square of the phonon frequency and is proportionate to $N(E\mathrm{_F})$). Therefore, it is meaningless to discuss phonon frequency alone. Moreover, the PdTe$_6$ octahedral unit is distorted after the SPT (inset of Fig. 4(c)), as such, the contraction of $\theta\mathrm{_{Te-Pd-Te}}$ (Fig, 3(g)) cannot be connected simplistically to the phonon softening in comparison with dilation before the SPT. Further experimental study of Raman scattering is needed to clarify the role of different phonon modes in the superconductivity of PdTe. Another way is to increase $N(E\mathrm{_F})$ to enhance the electron-phonon coupling directly. As mentioned previously, the normal-state (near superconducting state) electronic properties is a good indicator of $N(E\mathrm{_F})$, thus the practical $N(E\mathrm{_F})$ perhaps returns to growth as the increase of $T\mathrm{_c}$ in PdTe after the SPT. Directly supporting this observation, both electron and hole concentrations, along with mobilities, exhibit a significant increase in tandem with the enhancement of $T\mathrm{_c}$ under high pressures (Figs. 2(d)-(e)). Since the SPT, the superconducting mechanism in PdTe, especially under high pressure, where dramatic changes in carrier concentrations and mobilities occur, deserves further experimental and theoretical efforts.\par

In summary, we systematically investigated the pressure-induced effects on the electronic and structural properties of the Dirac semimetal PdTe. The minimum $T\mathrm{_c}$ of 1.9 K is identified at the critical pressure $P\mathrm{_c} \approx$ 15 GPa. Below $P\mathrm{_c}$, the $T\mathrm{_c}$ within the SC-\Rmnum{1} phase region behaves a monotonic decrease with increasing pressure, arising from the cooperation of density of states at the Fermi level $N(E\mathrm{_F})$ and phonon frequency. Upon surpassing $P\mathrm{_c}$, PdTe undergoes a structural transition into the SC-\Rmnum{2} phase region. In this phase, $T\mathrm{_c}$ experiences an increase from 1.9 K to 2.7 K before reaching a plateau of approximately 2.5 K as pressure varies. In this region, the increase in $T\mathrm{_c}$ is closely related to the carrier concentrations and mobilities. Both theoretical calculations and structural analysis consistently suggest that SC-\Rmnum{1} phase implements the framework of the BCS theory, while the SC-\Rmnum{2} phase requires further investigations to understand the origin of the unusual enhancement of SC. This work not only demonstrates the tunability of the superconducting phases of PdTe under pressure but also provides new insights to deepen our understanding of the mechanism of superconductivity in PdTe.

\begin{acknowledgments}
The authors acknowledge Xu Chen, Chunhua Xu and Boqin Song at Institute of Physics (IOP) of Chinese Academy of Sciences for fruitful discussion, Wei Xia at School of Physical Science and Technology at ShanghaiTech University for technical assistance in high-pressure synchrotron diffraction measurements. This work was supported by the Natural Science Foundation of China (Grant No. 12174064) and the Shanghai Municipal Science and Technology Major Project (Grant No. 2019SHZDZX01). Yanfeng Guo was supported by the open project of Beijing National Laboratory for Condensed Matter Physics (Grant No. ZBJ2106110017) and the Double First-Class Initiative Fund of ShanghaiTech University. Xiaofeng Xu was supported by the National Natural Science Foundation of China (Grants No. 12274369 and No. 11974061).
\end{acknowledgments}

\end{document}